\documentclass[%
 reprint,
 showpacs,preprintnumbers,
 amsmath,amssymb,
 aps,
 prb,
]{revtex4-1}
\usepackage{dcolumn}
\usepackage{subfigure}
\usepackage{graphicx,color}
\usepackage{mathrsfs}

\makeatletter
\def\btt#1{\texttt{\@backslashchar#1}}
\DeclareRobustCommand\bblash{\btt{\@backslashchar}} \makeatother

\begin{document}

\title{Dynamical conductivity in the multiply-degenerate point-nodal semimetal CoSi}

\author{Tetsuro Habe}
\affiliation{Department of Applied Physics, Hokkaido University, Sapporo, Hokkaido 060-0808, Japan}

\date{\today}

\begin{abstract}
We investigate the dynamical conductivity in multiply-degenerate point-nodal semimetal CoSi.
In the semimetal, the band structure holds point nodes at the $\Gamma$ and R points in the Brillouin zone and more than three bands touch at the nodes.
Around the nodes, electronic states are predicted to be described as the multifold chiral fermion, a new class of fermion.
We show that the dynamical conductivity exhibits a characteristic spectrum corresponding to the band structure and the chiral fermionic states.
The dynamical conductivity of CoSi is calculated as a function of photon energy by using the first-principles band calculation and linear response theory.
We show that a dip structure in the low photon-energy region is attributed to not only the band structure but also the chirality of electronic states.
The chirality leads to the prohibition of transition between the lower and upper bands of threefold chiral fermion and thus the transition between the middle and lower bands is relevant to the dynamical conductivity.
This transition property is different from the Dirac and Weyl semimetals, the other point-nodal semimetals, where the excitation between the upper and lower bands is relevant to the dynamical conductivity.
We discuss the relation between the prohibition and the dip structure by using an effective Hamiltonian describing threefold chiral fermion.
\end{abstract}

\maketitle
\section{Introduction}
Topological nodal semimetals have attracted much attention in condensed matter physics due to the electronic structure and unique phenomena associated with it.
In point-nodal semimetals, the conduction and valence bands touch each other at points, called nodes, in the Brillouin zone.\cite{Murakami2007}
Dirac and Weyl semimetals are the first generation of topological point-nodal semimetals.\cite{Wan2011,Yang2011,Burkov2011-weyl,Neupane2013,Borisenko2013}
In these semimetals, electronic states around the nodes are described by Dirac equation in high-energy physics and have provided a test ground for studying the property of Dirac and Weyl fermions in condensed matter.
Recently, a novel type of point-nodal semimetals has been proposed theoretically and predicted to have multiply degenerate electronic states, so-called multifold chiral fermion, which can not be described by Dirac equation.\cite{Wieder2016,Bradlyn2016,Bradlyn2017,Tang2017,Bradlyn2018}
The chiral fermion can emerge in some crystals as a robust electronic state topologically protected.\cite{Cheng2018}
Such topological nodal semimetals have been expected to provide novel electronic property beyond Dirac fermions.
For instance, the unique surface states\cite{Chang2017,Sanchez2019,Schroter2019} and the quantized photogalvanic response\cite{Chang2017,Flicker2018,Rees2019} have been studied theoretically and experimentally.

CoSi is a strong candidate of the novel topological semimetals according to the first principles band calculation.\cite{Tang2017,Cheng2017,Dutta2018,Severin2018}
Angle-resolved photoemission spectroscopy (ARPES) is one of the experimental techniques for identifying the band structure of topological semimetal.\cite{Neupane2013,Borisenko2013}
Recently, a few groups have reported the band structure of CoSi by using ARPES and shown nodes and topological surface states, one of the characteristic features of the topological semimetal.\cite{Takane2019,Rao2019}
However, there is a restriction in the visible electronic energy for ARPES because it is able to detect electronic states up to the Fermi level.
In the pristine crystal of the semimetal, the Fermi level is pined at the charge neutral point, and it is slightly under one of the nodes in CoSi.
%Therefore, the other measurement is necessary for revealing the electronic property of multi-fold chiral fermion above the Fermi level.

In this paper, we theoretically investigate the dynamical conductivity of CoSi by using the first-principles calculation and linear response theory.
The dynamical conductivity is associated with inter-band transitions between the conduction and valence bands, and thus it is able to provide the information of electronic bands above the Fermi level.
The photon energy dependence of dynamical conductivity is strongly affected by the band structure.
In conventional semiconductors, the spectrum exhibits a quadratic dependence on the photon energy.
However it shows a linear dependence in Weyl semimetals and can be flat, independent of photon energy, in nodal-line semietals. 
In Ref.\ \onlinecite{Habe2018}, the author and co-worker have shown that the numerical result by using the first-principles calculation agrees well with the experimental measurement for a topological nodal-line semimetal \cite{Schilling2017} even in a high photon-energy region.
We provide the realistic numerical result and discuss the relation between the spectrum of dynamical conductivity and the electronic band structure.
Moreover, we find that the spectrum below 250 meV is ascribable to not only the band structure but also the property of multifold chiral fermion; a prohibition of electronic transition between a specific pair of bands.
The multifold chiral fermion imposes a dip structure in the spectrum of dynamical conductivity around the photon energy of 200 meV.

%%%%%%%%%% FIGURE %%%%%%%%%%
\begin{figure}[htbp]
\begin{center}
 \includegraphics[width=80mm]{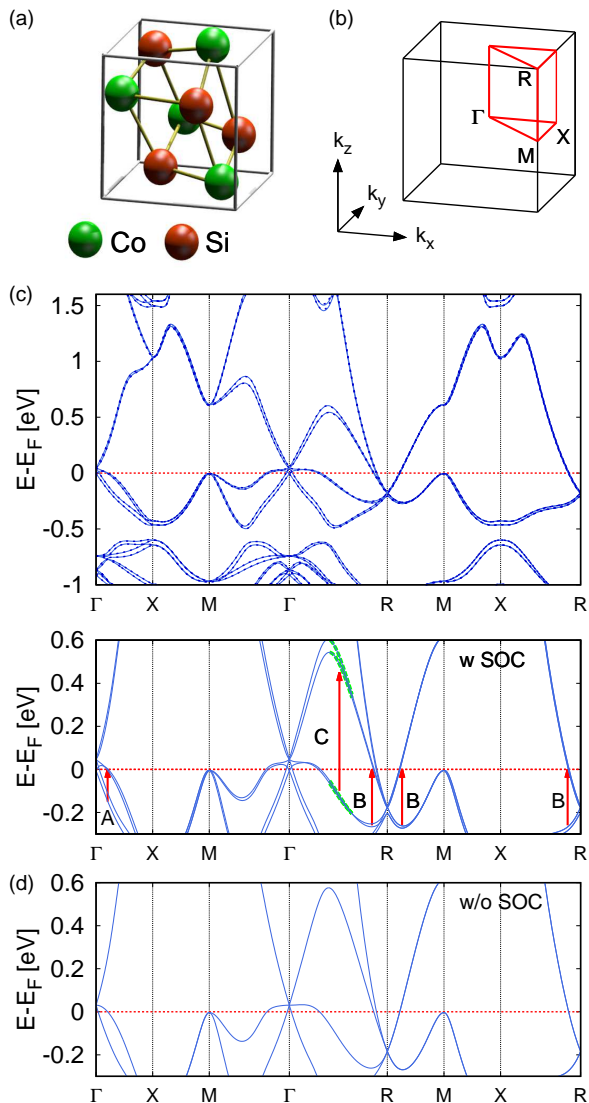}
\caption{The unit cell and the Brillouin zone for CoSi in (a) and (b), respectively. The band structure of CoSi calculated by the first-principles calculation including the spin-orbit coupling in (c). The band structure within different energy ranges is depicted in the upper and lower panels. The dashed line in the upper panel indicates the band calculated by using the tight-binding model discussed in Sec.\ \ref{sec_calculation}. In the lower panel, the arrows A, B, and C indicate the transition processes corresponding to the characteristic structure of dynamical conductivity in Fig.\ \ref{fig_optical_conductivity_temperature} (a). The highlighted region is associated with the first peak around 650 meV in Fig.\ \ref{fig_optical_conductivity_self_energy}.  In (d), the band is calculated without the spin-orbit interaction.
 }\label{fig_band_structure}
\end{center}
\end{figure}
%%%%%%%%%%%%%%%%%%%%%%%%%

%%%%%%%%%%%%%%%%%%%%
\section{First-principles band structure}
%%%%%%%%%%%%%%%%%%%%
We calculate the electronic band structure of CoSi by using quantum ESPRESSO, a first-principles calculation code,\cite{Quantum-espresso} and show it in Fig.\ \ref{fig_band_structure} (c).
The cubic unit cell and the Brillouin zone are schematically depicted in Fig.\ \ref{fig_band_structure} (a) and (b), respectively.
The lattice constant and the atomic position are computed within the same code by relaxing the lattice to the stable structure.
Then, the lattice constant is estimated to be 4.432{\AA} and the atomic position is given in Table\ \ref{table_atomic_position}.
The crystal structure is classified into $P2_13$ and represented by two parameters of $x_{\mathrm{Co}}=0.144$ and $x_{\mathrm{Si}}=0.843$ for Co and Si, respectively.
We apply a projector augmented wave method to the first-principles calculation with a generalized-gradient approximation functional including spin-orbit coupling (SOC).
The cut-off energy of plane wave basis and the convergence criterion are adopted 50 Ry and $10^{-8}$ Ry, respectively.
\begin{table}[ht]
\begin{center}
\begin{tabular}{c c c c}
\hline
\hline
&$x$&$y$&$z$\\ \hline
Co$^{(1)}$&0.144&0.144&0.144\\ 
Co$^{(2)}$&0.644&0.356&0.856\\ 
Co$^{(3)}$&0.356&0.856&0.644\\ 
Co$^{(4)}$&0.856&0.644&0.356\\ 
Si$^{(1)}$&0.843&0.843&0.843\\
Si$^{(2)}$&0.343&0.657&0.157\\
Si$^{(3)}$&0.657&0.157&0.343\\
Si$^{(4)}$&0.157&0.343&0.657\\ \hline \hline
\end{tabular}\label{table_atomic_position}
\end{center}
\caption{The list of atomic positions in the unit cell of CoSi, which is shown in Fig.\ \ref{fig_band_structure}(a). The positions are represented in units of the lattice constant.}
%\end{ruledtabular}
\end{table}

We show the first-principles band structure of CoSi in Fig.\ \ref{fig_band_structure}(c), where is is presented within the different energy ranges in the upper and lower panels.
Nodal-points appear at two high-symmetry points; the $\Gamma$ point and the R point.
At the $\Gamma$ point, electronic states are quadruply degenerated at the upper node and doubly degenerated at the lower node, where the slight energy-split between nodes is due to SOC\cite{Ishii2014}.
In the absence of SOC, a single node appears with triply-degenerated electronic states for each spin as shown in Fig.\ \ref{fig_band_structure}(d).
The fourfold degenerated node is slightly above the Fermi level, and a doubly degenerate node emerges under the level. 
At the R point, a single node appears with sextuple degeneration 175meV below the Fermi energy.
Four cones and a single quadratic band are crossing at the node.
In the band structure without SOC, there is a fourfold degenerate node emerges for each spin at the R point.
%The local maximum of valence band emerges at the $M$ point with a gap energy of 620meV.

%%%%%%%%%%%%%%%%%%%%
\section{Dynamical conductivity}\label{sec_calculation}
%%%%%%%%%%%%%%%%%%%%
The dynamical conductivity $\sigma(\omega)$ is the response function to the oscillating electric field with the frequency $\omega$ and is related to the optical property of materials.
The real part $\sigma_1(\omega)=\mathrm{Re}[\sigma(\omega)]$ is proportional to the optical absorption, and it is also associated with the reflectance, the imaginary part, by Kramers-Kronig analysis.
The photon-energy dependence shows the unique spectrum depending on the band structure of nodal semimetal; a flat spectrum for graphene,\cite{Ando2002,Gusynin2006,Koshino2008}, a linear dependence for Dirac and Weyl semimetals,\cite{Tabert2016,Mukherjee2017,Ahn2017}, and nearly flat spectrum for nodal-line semimetals\cite{Barati2017,Habe2018,Ebad2019}.
These spectra have been observed in some experimental fabrications.\cite{Stauber2008,Mak2008,Schilling2017}

In general, $\sigma_1(\omega)$ is associated with the electronic transition corresponding to the frequency of applied electric field, i.e., the photon energy.
We represent the electronic states in CoSi by using a multi-orbital tight-binding model defined on the basis of Wannier orbitals.
We adopt five $d$-orbitals in Co and three $p$-orbitals in Si as the basis and compute the spin-dependent hopping integrals and on-site potential from the first-principle bands in Fig.\ \ref{fig_band_structure} by using Wannier90, a code for calculating maximally localized Wannier functions and the hopping matrix\cite{Wannier90}.
This tight-binding model reproduces the first-principles band structure as shown in Fig.\ \ref{fig_band_structure}.

%%%%%%%%%% FIGURE %%%%%%%%%%
\begin{figure}[htbp]
\begin{center}
 \includegraphics[width=75mm]{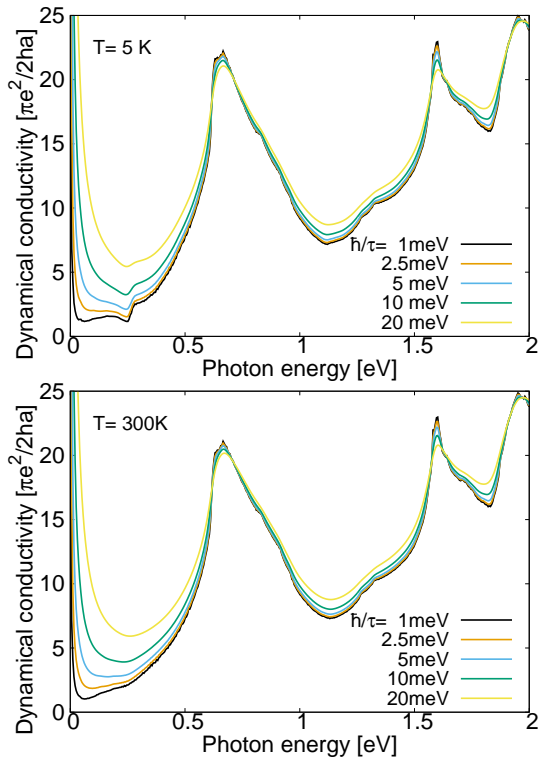}
\caption{The dynamical conductivity as a function of photon energy plotted for several relaxation-time, $\hbar/\tau=$1, 2.5, 5, 10, and 20 meV, at 5K in the upper panel, and that at 300K in the lower panel.
 }\label{fig_optical_conductivity_self_energy}
\end{center}
\end{figure}
%%%%%%%%%%%%%%%%%%%%%%%%%
We calculate the dynamical conductivity in the linear response theory where it for a linearly polarizing photon is represented by
\begin{align}
\sigma(\omega,T)=&\frac{2i}{\omega}\frac{e^2}{\hbar}\sum_{m\leq n}\int_{\mathrm{BZ}}\frac{d^3\boldsymbol{k}}{(2\pi)^3}\frac{\left|\langle n \boldsymbol{k}|\hat{v}_\alpha|m\boldsymbol{k}\rangle\right|^2}{\hbar\omega-(E_{n\boldsymbol{k}}-E_{m\boldsymbol{k}})+i\hbar/(2\tau)}\nonumber\\
&\times\left(n_F(E_{n\boldsymbol{k}},T)-n_F(E_{m\boldsymbol{k}},T)\right),\label{eq_dynamical_conductivity}
\end{align}
where $n_F(E,T)$ is Fermi distribution function at the temperature $T$ and the energy $E$ with respect to the Fermi level.
Here, $\omega$ is the frequency of photon, and the velocity is defined by the commutation of the position and the Hamiltonian, $\hat{v}_\alpha=(1/i\hbar)[x_\alpha,H]$, where $\alpha$ indicates parallel direction to the polarizing direction of photon.
The eigenstate $|m\boldsymbol{k}\rangle$ and the energy $E_{m\boldsymbol{k}}$ are calculated by using the multi-orbital tight-binding Hamiltonian.
The effect of impurity is included in Eq.\ \ref{eq_dynamical_conductivity} as the relaxation time $\tau$.

We consider the dynamical conductivity for linearly-polarized photons in (100) direction and show the relaxation time-dependence of $\sigma_1(\omega)$ at $T=5$K and 300K in Fig.\ \ref{fig_optical_conductivity_self_energy} .
The low-frequency spectrum is sensitive to $\tau$ in $\omega<500$meV because the tail of Drude peak around $\omega=0$ spreads with the increase in the impurity density, i.e., the decrease of relaxation time $\tau$.
When the material is clean, the inter-band transition is relevant to the spectrum in the low-frequency region.
%In $\hbar/\tau\leq5$ meV, the spectrum reflects the excitation between electronic states around nodes in the Brillouin zone.
In the nearly ballistic condition, $\hbar/\tau=1$meV, the dynamical conductivity linearly increases with $\omega$ up to $\omega\sim150$meV at both the temperatures.
The linear dependence is a characteristic feature of three-dimensional point-nodal semietals and attributed to the density of state.\cite{Tabert2016,Sanchez2019}
This characteristic structure disappears with the increase in $\hbar/\tau$, i.e., the increase of disorder, due to the broadening of Drude peak but it is stable in a clean sample even at the room temperature.

\begin{figure}[htbp]
\begin{center}
 \includegraphics[width=75mm]{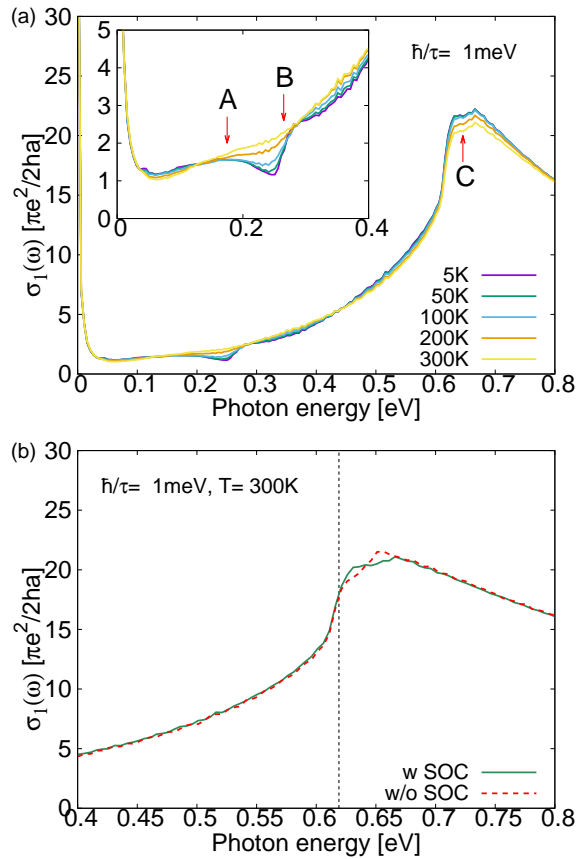}
\caption{The dynamical conductivity plotted at several temperatures, $T=$5, 10, 50, 100, and 300K, for $\hbar/\tau=$1meV and 10meV in (a). A, B, and C correspond to the excitation energies in the transition processes indicated in Fig.\ \ref{fig_band_structure} (c). In (b), the dynamical conductivity calculated with and without the spin-orbit coupling. The vertical dashed line indicates the minimal excitation energy at the M point. 
 }\label{fig_optical_conductivity_temperature}
\end{center}
\end{figure}

In Fig.\ \ref{fig_optical_conductivity_temperature}, we show the temperature dependence of $\sigma_1(\omega)$ up to the room temperature with $\hbar/\tau=$ 1meV in a narrower energy window.
%%%%%%%%%% FIGURE %%%%%%%%%%
%%%%%%%%%%%%%%%%%%%%%%%%%
The numerical result exhibits that the linear slope is unchanged with the increase in temperature.
In what follows, we investigate the structures of spectrum in this energy region and characterize them by the electronic transition in the band structure.

Firstly, we find a dip structure emerging around $\hbar\omega\sim250$ meV in the spectrum of $\sigma_1(\omega)$ below $200$K.
The step-like upward slope and the downward slop are attributed to different chiral fermions at the $\Gamma$ point and the R point in the Brillouin zone.
This dip structure is different from a similar structure associated with a single chiral fermion in a recent work\cite{Sanchez2019}. 
The dip structure in the previous work appears at the photon frequency corresponding to the energy of node with respect to the Fermi level, $\hbar\omega\sim10$ meV.
In the present work, however, the dip structure appears at much larger photon energy, $\hbar\omega<100$ meV.
In what follows, we discuss the upward and downward slopes of the dip structures separately.

The upward slope emerges with a step-like structure at the photon energy B in Fig.\ \ref{fig_optical_conductivity_temperature} (a).
The photon energy is corresponding to the excitation energy to electronic states in the Fermi level around the R point, where the transition is depicted as B in Fig.\ \ref{fig_band_structure}).
In general, the dynamical conductivity is a linear function of photon energy in the case of 
three-dimensional linear dispersion with the Fermi level crossing the node.\cite{Tabert2016}
When the upper (lower) cone is partially filled, there is no transition to electronic states, which are occupied, in the upper (lower) cone below (above) the Fermi level.
Thus the dynamical conductivity is zero below the photon energy corresponding to the transition to the Fermi level and shows a step-like upward slope at the energy.\cite{Mukherjee2017-2} 
Therefore the upward slope is attributed to the partially filled multifold-cone at the R point.

The downward slope and the lower spectrum, on the other hand, are ascribable to the multifold chiral fermion at the $\Gamma$ point. 
Since the excitation energy is smaller than that of B, the multifold cone at the R point has no contribution.
To confirm the relation to the chiral fermions, we consider an effective Hamiltonian describing electronic states around the $\Gamma$ point and compare the dynamical conductivity in two models.
Here we omit the SOC in the effective Hamiltonian because the low-energy dynamical conductivity is nearly unchanged with the presence of SOC.
The effective Hamiltonian is represented by a $3\times3$ matrix on the basis of three degenerated electronic states $\Psi_j$ for $j=$1-3 at the $\Gamma$ point.
The constant element gives the energy of node at the $\Gamma$ point and thus it is proportional the identity matrix.
We consider only matrix element linearly depending on the wave number $k_\mu$ in the effective model.
Such $k_\mu$-linear element can be represented by $\{H_{3\times3}\}_{ij}=\sum_\mu k_\mu\langle\Psi_i|V_\mu|\Psi_j\rangle$, where $V_\mu$ is defined by the derivative of the full Hamiltonian $\hbar^{-1}\partial H/\partial k_\mu$ and $|\Psi_j\rangle$ is the wave function at the $\Gamma$ point.
Here $V_\mu$ and $|\Psi_j\rangle$ are calculated numerically by using the Fourier transformed multi-orbital tight-binding Hamiltonian.

\begin{figure}[htbp]
\begin{center}
 \includegraphics[width=75mm]{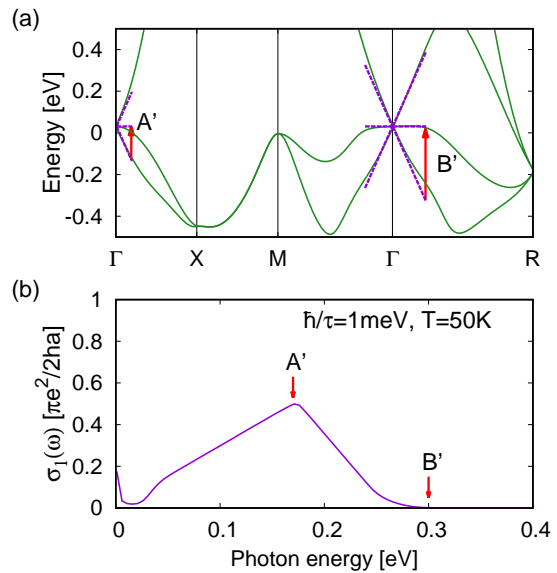}
\caption{The band structure (a) and the dynamical conductivity (b) calculated by using the effective model. The dashed line indicates the band structure calculated by using $k$-linear effective Hamiltonian. The Drude peak is eliminated in the calculation.
 }\label{fig_eff_band}
\end{center}
\end{figure}

\begin{figure}[htbp]
\begin{center}
 \includegraphics[width=75mm]{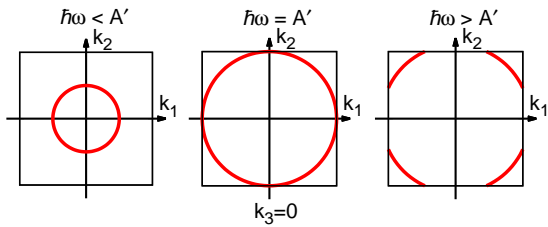}
\caption{The distribution of electronic states corresponding to the relevant transition for the dynamical conductivity in the effective model. Three panels schematically represent the case of different photon energy $\hbar\omega$. The wave number $(k_1,k_2,k_3)$ are any permutation of $(k_x,k_y,k_z)$.
 }\label{fig_transition_surface}
\end{center}
\end{figure}

We show the band structure calculated by using the effective Hamiltonian in Fig.\ \ref{fig_eff_band}(a).
The wave vector is confined in a cubic region $k_\alpha\in[-2\pi/(10a),2\pi/(10a)]$ where the cubic shape is adopted due to the experimentally observation of constant-energy surface.\cite{Takane2019}
The dynamical conductivity is also calculated by using the effective Hamiltonian and also shown at $\hbar/\tau=1meV$ and $T=50K$ in Fig.\ \ref{fig_eff_band}(b).
The structure of spectrum qualitatively agrees with that in Fig.\ \ref{fig_optical_conductivity_temperature} (a).
The spectrum has a peak at $\omega=170$meV (A') and vanishes around 300meV (B').
In Fig.\ \ref{fig_eff_band}, we indicate two transition processes corresponding to A' and B'.
Both transitions occur at the edge of the cubic region and the excitation energy is minimum (maximum) at A' (B') in the edge.
In Fig.\ \ref{fig_transition_surface}, we schematically depict the distribution of electronic states associated with the resonant transition between the middle and lower bands to the dynamical conductivity for some photon energies $\hbar\omega$.
The electronic states isotropically distributed in the cubic region for a photon energy lower than the excitation energy of A'.
When $\hbar\omega$ exceed the excitation energy of A', the resonant transition is absent between the middle and lower bands in some direction of wave vector and disappears above the excitation energy B'.
The dynamical conductivity also decreases to zero between A' and B' energies even though the larger excitation energy is present between the lower and upper bands.
Thus this result implies the prohibition of transition between the lower and upper bands.
This prohibition is a unique feature of the chiral fermion and essentially different from the case of Dirac fermion, where the inter-cone excitation is relevant to the dynamical conductivity\cite{Ando2002}.
In the next section, we confirm the prohibition for threefold chiral fermions analytically.
%%%%%%%%%% FIGURE %%%%%%%%%%
%%%%%%%%%%%%%%%%%%%%%%%%%

Secondly, we focus on a peak with a flat-top around $\omega\simeq650$meV.
This flat-top peak is attributed to the parallel structure of bands and exhibits the effect of SOC to the nodal band structure around the R point.
We represent the electronic states associated with the relevant transition to the dynamical conductivity as a shaded line in Fig.\ \ref{fig_band_structure}(c).
The occupied and unoccupied states are distributed in the aligned lines of band structure.
In the aligned occupied and unoccupied bands, electronic states show the resonant transition at the nearly same photon energy and produce a peak structure similar in the case of the nodal-line semimetal\cite{Habe2018}.
Moreover, the spin-split of upper band affects the structure of peak.
In the practice, the peak structure changes with the presence of SOC as shown in Fig.\ \ref{fig_optical_conductivity_temperature}(b).
In general, the spin-split of band leads to the split of peak in the spectrum of dynamical conductivity because the spin-split leads to two resonant states.
However, the spin-split in the parallel bands does not split the peak structure.
Since the split gradually decreases to zero with the wave number toward the R point, electronic states have different excitation energies distributed in a range corresponding to the SOC split and they produce a flat-top peak consisting of a series of peaks with the energy continuously changing.
Therefore, the flat-top peak exhibits the SOC split in the band of multifold chiral fermion at the R point.
%%%%%%%%%% FIGURE %%%%%%%%%%
%%%%%%%%%%%%%%%%%%%%%%%%%

Finally, in higher-frequency region, we find the other peaks at $\omega\simeq1.6$ eV and 1.9 eV.
The two peaks are attributed to the conventional mechanism based on the local maximum and minimum of energy dispersion at high symmetry points.
In the local maximum and minimum, the density of states is enhanced and thus large number of states contribute to the dynamical conductivity.
The peak at 1.6 eV is corresponding to the excitation between the lowest conduction band and the second highest valence band around -1 eV at the M point in Fig.\ \ref{fig_band_structure}(c).
The another peak emerges at 1.9 eV due to the excitation between the highest valence band and the band slightly above 1.5 eV at the X point.
At these points, the transition between the lowest conduction band and the highest valence band.
This is because the electronic states have the same parity under twofold rotation.
These high-symmetry points, the M and X points, are unchanged under twofold rotation within the first Brillouin zone.
Since the twofold rotation along a direction changes the sign of velocity in two orthogonal directions, the transition is prohibited between two electronic states with the same parity under the twofold rotation due to the absence of matrix element of velocity between them.
Therefore, the transition between the closest occupied and unoccupied states is absent at the M and X points and has no contribution to the dynamical conductivity.

%%%%%%%%%%%%%%%%%%%%%%%%%%%%
\section{Effective theory}
%%%%%%%%%%%%%%%%%%%%%%%%%%%%
In this section, we analyze the electronic transition probability in the low excitation energy region by using an effective model based on the crystal symmetry.
We ignore the effect of spin-orbit coupling in this analysis because the coupling does not have a relevant contribution to the dynamical conductivity in this energy region according to the numerical calculation.

The low energy transition occurs between electronic states around the $\Gamma$ point.
Three electronic states are degenerate at the high symmetry point.
Since CoSi is classified into $P2_13$, the degenerated states at the $\Gamma$ point are bases of an irreducible representation $T$.
The crystal structure has three axes of twofold skew symmetry; (100), (010), and (001), and four axes of threefold rotation symmetry; $(111)$, $(1\bar{1}\bar{1})$, $(\bar{1}\bar{1}1)$, and $(\bar{1}1\bar{1})$.
The index of the representation is $\chi=3$ for identity operation, $\chi=-1$ for twofold skew rotation, and $\chi=0$ for threefold rotation.
In what follows, we consider wave functions consisting of wannier orbitals in the representation $T$.

The first principles calculation reveals that the  the $d$-orbitals in Co atoms constitute a significant part of the electronic states around the Fermi level.
In the unit cell, there are four Co atoms at $\boldsymbol{d}_j$, where the atomic positions are given by
\begin{align}
\begin{split}
\boldsymbol{d}^{(1)}&=(x_{\mathrm{Co}}+1/4,x_{\mathrm{Co}}-1/2,x_{\mathrm{Co}})\\
\boldsymbol{d}^{(2)}&=(x_{\mathrm{Co}}-1/4,-x_{\mathrm{Co}},-x_{\mathrm{Co}})\\
\boldsymbol{d}^{(3)}&=(-x_{\mathrm{Co}}-1/4,-x_{\mathrm{Co}}+1/2,-1/2+x_{\mathrm{Co}})\\
\boldsymbol{d}^{(1)}&=(-x_{\mathrm{Co}}+1/4,x_{\mathrm{Co}},1/2-x_{\mathrm{Co}}).
\end{split}
\end{align}
Here we shift the origin for the clear explanation about the symmetry in comparison with Table \ref{table_atomic_position}.
Then the wave function can be represented by a superposition of wannier orbitals at the four positions.
Twofold skew rotation about $(100)$-direction exchanges these atoms as $(\boldsymbol{d}^{(1)},\boldsymbol{d}^{(2)},\boldsymbol{d}^{(3)},\boldsymbol{d}^{(4)})\rightarrow(\boldsymbol{d}^{(3)},\boldsymbol{d}^{(4)},\boldsymbol{d}^{(1)},\boldsymbol{d}^{(2)})$.
Threefold rotation about $(111)$  leads to $(\boldsymbol{d}^{(1)},\boldsymbol{d}^{(2)},\boldsymbol{d}^{(3)},\boldsymbol{d}^{(4)})\rightarrow(\boldsymbol{d}^{(1)},\boldsymbol{d}^{(3)},\boldsymbol{d}^{(4)},\boldsymbol{d}^{(2)})$ with the rotation angle $\theta=2\pi/3$.
Then we can adopt three degenerated states at the $\Gamma$ point as
\begin{align}
\begin{split}
\Psi_1(\boldsymbol{r})=&\frac{1}{2}\{\psi_m^{(111)}(\boldsymbol{r}-\boldsymbol{d}^{(1)})-\psi_m^{(1\bar{1}\bar{1})}(\boldsymbol{r}-\boldsymbol{d}^{(2)})\\
&+\psi_m^{(\bar{1}\bar{1}1)}(\boldsymbol{r}-\boldsymbol{d}^{(3)})-\psi_m^{(\bar{1}1\bar{1})}(\boldsymbol{r}-\boldsymbol{d}^{(4)})\}\\
\Psi_2(\boldsymbol{r})=&\frac{1}{2}\{\psi_m^{(111)}(\boldsymbol{r}-\boldsymbol{d}^{(1)})+\psi_m^{(1\bar{1}\bar{1})}(\boldsymbol{r}-\boldsymbol{d}^{(2)})\\
&-\psi_m^{(\bar{1}\bar{1}1)}(\boldsymbol{r}-\boldsymbol{d}^{(3)})-\psi_m^{(\bar{1}1\bar{1})}(\boldsymbol{r}-\boldsymbol{d}^{(4)})\}\\
\Psi_3(\boldsymbol{r})=&\frac{1}{2}\{\psi_m^{(111)}(\boldsymbol{r}-\boldsymbol{d}^{(1)})-\psi_m^{(1\bar{1}\bar{1})}(\boldsymbol{r}-\boldsymbol{d}^{(2)})\\
&-\psi_m^{(\bar{1}\bar{1}1)}(\boldsymbol{r}-\boldsymbol{d}^{(3)})+\psi_m^{(\bar{1}1\bar{1})}(\boldsymbol{r}-\boldsymbol{d}^{(4)})\},
\end{split}\label{eq_basis}
\end{align}
where $\psi_m^\nu(\boldsymbol{r})$ is the $d$-orbital function with the character of $m$ and the polar axis $\nu$, and it is real function due to time-reversal symmetry.
In this basis, the representations of twofold and threefold rotations; $R^\nu_{1/2}$ and $R^\nu_{n/3}$, respectively,  are given by
\begin{align}
R_{1/2}^{(100)}=&\begin{pmatrix}
-1&0&0\\
0&1&0\\
0&0&-1
\end{pmatrix},\\
R_{1/3}^{(111)}=&e^{i(2\pi/3)m}\begin{pmatrix}
0&0&1\\
1&0&0\\
0&1&0
\end{pmatrix}=\{R_{2/3}^{(111)}\}^{-1}.\label{eq_representations}
\end{align}
The representations for the other operations can be obtained as a matrix product of $R_{1/2}^{(100)}$ and $R_{1/3}^{(111)}$.

We discuss an effective Hamiltonian describing electronic states around the $\Gamma$ point by using the representations.
Three functions $\Psi_j$ are degenerated at the $\Gamma$ point but they split into three at non-zero wave numbers.
In the vicinity of the $\Gamma$ point, electronic states can be described by the $k$-linear Hamiltonian
\begin{align}
H^{(1)}_{\mathrm{eff}}=\hat{v}_xk_x+\hat{v}_yk_y+\hat{v}_zk_z,
\end{align}
where $\hat{v}_\mu$ is a $3\times3$ velocity matrix on the bases of $\Psi_j$ and we set the node to be the origin of energy.
When we consider electronic states in the $k_\mu$ axis, the velocity matrix $\hat{v}_\mu$ represents the Hamiltonian, $H^{(1)}_{\mathrm{eff}}=\hat{v}_\mu k_\mu$, and thus it is invariant under $R_{1/2}^{\mu}$.
Therefore, for instance, $\hat{v}_x$ can be represented by
\begin{align}
\hat{v}_x=&
v_{0x}
\begin{pmatrix}
1&0&0\\
0&0&0\\
0&0&1
\end{pmatrix}
+
v_{1x}
\begin{pmatrix}
0&0&1\\
0&0&0\\
1&0&0
\end{pmatrix}
+
v_{2x}
\begin{pmatrix}
0&0&-i\\
0&0&0\\
i&0&0
\end{pmatrix}\nonumber\\
&+
v_{3x}
\begin{pmatrix}
1&0&0\\
0&0&0\\
0&0&-1
\end{pmatrix}
+
v_{4x}
\begin{pmatrix}
0&0&0\\
0&1&0\\
0&0&0
\end{pmatrix}.
\end{align}
In the pristine CoSi, time-reversal symmetry is preserved besides crystal symmetries.
Since all bases are unchanged under time-reversal operation, time-reversal operation for the Hamiltonian is $\mathcal{T}H(\boldsymbol{k})\mathcal{T}^{-1}=H^\ast(-\boldsymbol{k})$.
Thus time-reversal symmetry requires $v_{jx}$ to be non-zero only for $j=2$, i.e., $v_{jx}$ must be real.
Moreover the velocity operators in the other two directions are obtained by threefold rotation in Eq.\ (\ref{eq_representations}).
Therefore the effective Hamiltonian is written by
\begin{align}
H_{\mathrm{eff}}^{(1)}=iv\begin{pmatrix}
0&-k_y&k_x\\
k_y&0&-k_z\\
-k_x&k_z&0
\end{pmatrix}.
\end{align}
Here the Hamiltonian is obtained by setting $\theta=\pi/2$ in the general $k$-linear Hamiltonian for threefold degenerate fermion\cite{Bradlyn2017} and equivalent to a model proposed in Ref. \onlinecite{Manes2012}. 
By using the data of first-principle band, the velocity coefficient is estimated to be $v\sim 6.7\times10^7$ cm/s around the $\Gamma$ point.
The vector component of Bloch functions are given by
\begin{align}
\psi_0=\frac{1}{k}\begin{pmatrix}
k_z\\
k_x\\
k_y
\end{pmatrix},\;\;
\psi_\pm=\frac{1}{k\sqrt{2(k_x^2+k_y^2)}}
\begin{pmatrix}
k_x^2+k_y^2\\
\pm ikk_y-k_xk_z\\
\mp ikk_x-k_yk_z
\end{pmatrix},
\end{align}
with the energy $E=0$ and $\pm vk$, respectively. 
The velocity operators in the $x$, $y$, and $z$ directions map $\psi_-$ onto $\psi_0$ and $\psi_-$ as
\begin{align}
\langle \psi_0|
\begin{pmatrix}
\hat{v}_x\\
\hat{v}_y\\
\hat{v}_x
\end{pmatrix}
|\psi_-\rangle
=&
\frac{v}{k\sqrt{2(k_x^2+k_y^2)}}
\begin{pmatrix}
ikk_y+k_xk_z\\
ikk_x-k_yk_z\\
k_x^2+k_y^2
\end{pmatrix}
\\
\langle \psi_-|
\begin{pmatrix}
\hat{v}_x\\
\hat{v}_y\\
\hat{v}_z
\end{pmatrix}
|\psi_-\rangle
=&
\frac{v}{k}
\begin{pmatrix}
k_x\\
-k_y\\
-k_z
\end{pmatrix}.
\end{align}
Any velocity operator has zero as matrix elements between $\psi_-$ and $\psi_+$.
Therefore the transition between the lower band and the upper band is prohibited, and thus it has no contribution to the dynamical conductivity and leads to the characteristic structure in Fig.\ \ref{fig_optical_conductivity_temperature} (a) and \ref{fig_eff_band} (b).

The prohibition is a characteristic feature of time-reversal symmetric threefold chiral fermion in comparison with Weyl and Dirac semimetals, the other systems with the linear dispersion.
Threefold chiral fermion is described by a purely imaginary Hamiltonian in the presence of time-reversal symmetry.
Thus the wave functions in the upper and lower bands are transformed by taking complex conjugate, $\psi_+=\psi_-^\ast$.
The velocity operator is also purely imaginary and thus its matrix element satisfies $\langle \psi_+|\hat{v}_\mu|\psi_-\rangle^\ast=-\langle \psi_-|\hat{v}_\mu|\psi_+\rangle=0$, where the last equality is due to the hermitian property of velocity operator.
Therefore the transition between the upper and lower bands has no contribution to the dynamical conductivity in Eq.\ (\ref{eq_dynamical_conductivity}).
Dirac and Weyl semimetals, on the other hand, show transitions between the upper and lower bands even in the presence of time-reversal symmetry.
This is because the electronic states are described by Dirac Hamiltonian, which is not purely imaginary.
There are even number of point-nodes including degenerate nodes, where the nodes are divided into time-reversal pairs.
The time-reversal partner allows the presence of real part in the Hamiltonian and the transition between the upper and lower bands.

Finally, we discuss the $k$-quadratic terms for the effective Hamiltonian.
According to the multi-orbital tight-binding model, the relevant term is represented by the identity matrix in the three-state basis in Eq.\ (\ref{eq_basis}).
Thus the prohibition between the upper and lower bands is preserved even in the presence of the $k$-quadratic terms.
In the first-principle band, the middle band drops and crosses over the Fermi level with the increase in $k$.
This behavior enhances the excitation energy between the middle and upper bands and prohibits the transition between the middle and lower bands, i.e., the transition between two occupied states.
The prohibition due to the chiral nature of threefold fermions produces the reduction of dynamical conductivity at A  in Fig.\ \ref{fig_optical_conductivity_temperature}(a) and leads to the dip structure.

\section{Conclusion}
We have investigated the dynamical conductivity of CoSi, a novel topological semimetal holding multifold chiral fermion, by using first-principle band calculation and linear response theory.
The spectrum of dynamical conductivity $\sigma_1(\omega)$ exhibits the dip and peak structures at some photon frequency $\omega$.
We investigate the relation between these structures and the electronic structure.
The peak structures in $\hbar\omega$>1eV are corresponding to the transition between the local maximum and minimum of the occupied and unoccupied bands, respectively.
The peak structure at $\hbar\omega=650$meV, on the other hand, is attributed to the electronic transition between parallel bands and shows the flat-top structure due to the spin-split.

In the lower photon energy region, the spectrum reflects the property of multifold chiral fermions.
The dynamical conductivity linearly increases with the photon energy and turns down at $\hbar\omega=170$meV.
This behavior is attributed to not only the band structure but also the chirality of multifold fermion at the $\Gamma$ point.
We show the prohibition of transition between the upper and lower bands of threefold chiral fermion and that the prohibition leads to the decrease in the dynamical conductivity.
Then the partially-filled linear dispersion at the R point contributes to the increase again at $\hbar\omega=$200meV.
The two multifold chiral fermionic states at the $\Gamma$ and R points co-produce a dip structure in the spectrum of dynamical conductivity of CoSi.

\bibliography{CoSi}
\end{document}